%% file: paper.tex
\documentclass[12pt,a4paper,dvips]{article}
\usepackage{a4p}
\usepackage{cite,mcite}
\usepackage{graphicx}
\usepackage{rotating}
\usepackage{physics}
\usepackage{l3_title,ifthen,Lep}
\journalname{Phys. Lett. B}
\preprint{2002-081}
\date{November 5, 2002}

\newlength{\capindent}
\newlength{\capwidth}
\newlength{\figwidth}
\newcommand{\icaption}[2][!*!,!]{\hspace*{\capindent}%
  \begin{minipage}{\capwidth}
    \ifthenelse{\equal{#1}{!*!,!}}%
      {\caption{#2}}%
      {\caption[#1]{#2}}
  \end{minipage}}

\newcommand{\GG}{\ensuremath{\gamma ^* \gamma ^*}}
\newcommand{\Gg}{\ensuremath{\gamma \gamma}}
\newcommand{\pz}{\ensuremath{\pi^0}}
\newcommand{\ks}{\ensuremath{\rm K^0_S}}
\newcommand{\ppm}{\ensuremath{\pi^{\pm}}}
\newcommand{\kpm}{\ensuremath{\rm{K}^{\pm}}}

\newcommand{\Wgg}{\ensuremath{W_{\Gg}}}

\newcommand{\sqs}{\ensuremath{\sqrt{s}}}

\newcommand{\dpt} {\ensuremath{d\sigma / dp_t}}
\newcommand{\deta} {\ensuremath{d\sigma / d|\eta|}}

\begin{document}
      
\begin{titlepage}
\title{Inclusive Charged Hadron Production \\ 
in Two-Photon Collisions at LEP}
\author{The L3 Collaboration}

\begin{abstract}
Inclusive charged hadron production, 
$\rm{e}^{+} \rm{e}^{-} \rightarrow \rm{e}^{+} \rm{e}^{-}$ h$^{\pm}$ X,
is studied 
using 414 \pb {} of data collected at LEP with the L3 detector  
at centre-of-mass energies between 189 and 202 \GeV . 
Single particle inclusive differential cross sections
are measured as a function of the particle transverse momentum, \pt ,
and pseudo-rapidity, $\eta$.
For $ \pt \le 1.5 \GeV $, the data are well 
described by an exponential, typical of soft hadronic processes.
For higher \pt ,  the onset of 
perturbative QCD processes is observed.
The \pipm {} production cross section for $\pt > 5 \GeV$ is much higher than
the NLO QCD predictions.
\end{abstract}

\submitted 

\end{titlepage}

\section{Introduction}

Two-photon collisions are the main source of hadron production 
in the high-energy regime of LEP 
 via the process ${\rm e}^{+} {\rm e}^{-} \rightarrow {\rm e}^{+} {\rm e}^{-}
 \gamma ^{*}  \gamma ^{*}  \rightarrow 
 {\rm e}^{+} {\rm e}^{-}  hadrons$.
In the Vector Dominance Model (VDM), each photon can transform
into a vector meson with the same quantum numbers, 
thus initiating a strong interaction process with characteristics 
similar to hadron-hadron interactions. This
process dominates in the ``soft'' interaction region, where hadrons are produced
  with a low  transverse momentum, \pt , with respect to the beam direction.
Hadrons with  high \pt {} are produced by the direct QED process $\GG \ra \qqbar$ 
or by QCD processes originating from the partonic content of the photon.
QCD calculations are available 
for single particle inclusive production in two-photon interactions at 
next-to-leading order (NLO)  precision \cite{gordon1,kniehl}.

\par
The L3 collaboration recently published results on inclusive \pz {} and \ks {} 
production \cite{pz}. The \pz {} differential cross section
measured as a function of \pt {}
exhibits a clear excess over QCD calculations.
A comparison of these results with other single particle inclusive production
at high \pt {} is therefore important.
In this Letter, the inclusive charged hadron production is studied for a  
centre-of-mass energy  of the two interacting photons, \Wgg ,  
greater than 5 \GeV. The hadrons are measured 
in the transverse momentum range \mbox{$0.4 \GeV \le \pt \le 20 \GeV$} and in 
the pseudo-rapidity\footnote{$\eta  =  -\ln ~ \tan (\theta / 2)$, where 
$\theta$ is the polar angle of the particle relative to the beam axis.}
interval $|\eta| \le 1$.
The contributions from \pipm {} and \kpm {} are also derived.

\par 
The data used for this analysis were collected by the L3 detector \cite{L3}
 at centre-of-mass energies \sqs {} = 189 $-$ 202 \GeV , with a
luminosity weighted  average 
value of \sqs {} = 194 \GeV, for an
 integrated luminosity of 414 \pb .
Results on inclusive charged hadron production for a smaller data sample
at lower \sqs {} were previously reported \cite{opal}.

\par
The  process ${\rm e}^{+} {\rm e}^{-} \rightarrow {\rm e}^{+} {\rm e}^{-}   hadrons$  
is modelled with the PYTHIA  \cite{PYTHIA} event generator 
for an event sample three times larger than the data.
In this generator, each photon can interact as a point-like particle, as a vector meson or
as a resolved photon, leading to six classes of events. 
The fragmentation is simulated with JETSET.
Predictions from the PHOJET Monte Carlo program \cite{Engel} are also compared with the data.
The following Monte Carlo generators are used to simulate the relevant 
background processes: KK2f\cite{KK2f} for 
\ee $\rightarrow \rm q \bar{q} \,(\gamma $); 
KORALZ \cite{KORALZ} for \ee $\rightarrow \tau^{+} \tau^{-}(\gamma )$;
KORALW \cite{KORALW} for \ee $\rightarrow \rm{W}^{+} \rm{W}^{-}$ 
and  DIAG36 \cite{DIAG36} for \ee \ra {} \ee $\tau^{+} \tau^{-}$.
Events are simulated in the L3 detector using the GEANT \cite{GEANT}
and GHEISHA \cite{GHEISHA} programs
and passed through the same reconstruction program as the data.
Time dependent detector inefficiencies, as monitored during each data taking
period, are also simulated.

\section{Event and charged hadron selection}

\par
Two-photon  events  are collected predominantly by the track triggers \cite{tracktrig}
with a low \pt {} threshold of about 150 \MeV .
The selection of  ${\rm e}^{+} {\rm e}^{-} \rightarrow 
{\rm e}^{+} {\rm e}^{-} hadrons$ events \cite{l3tot} consists of:

\begin{itemize}
\item 
A multiplicity cut. To select hadronic final states, at  
least six objects must be detected,
where an object can be a track or a
calorimetric cluster with no associated track. 

\item 
Energy cuts. The total  energy deposited in the calorimeters must be 
less than 0.4 \sqs ,
in order to exclude \ee {} annihilation events.
The total energy in the electromagnetic calorimeter is required to be 
greater than 500 \MeV , to suppress
 beam-gas and beam-wall backgrounds.

\item
An anti-tag condition. 
Events with a cluster in  the luminosity monitor
with an energy greater than 30 \GeV
{} and an electromagnetic shower shape are excluded. 

\item
A mass cut. The visible mass of the event must be greater than 5 \GeV .

\end{itemize}

About 2 million  hadronic events are selected by these criteria.
The overall background level is less than  1\% and is 
mainly due to the $\ee \rightarrow \rm q \bar{q} \,(\gamma)$ and
 \ee \ra {} \ee $\tau^{+} \tau^{-}$  processes.

\par
Charged hadrons are measured with high quality tracks in the 
inner tracking detector. 
These tracks have a
transverse momentum greater than 400 \MeV {} and a distance of closest 
approach  to the primary vertex in the transverse plane less than 4 mm. 
The number of hits must be
greater than 80\% of that expected from the track length. 
Tracks are analysed in the $|\eta| < 1$ and $\pt < 20 \GeV$ range
where the detector resolution is optimal. A resolution $\sigma_{\pt} / \pt 
\simeq 0.015(\GeV^{-1})\times\pt$ is achieved.

\section{Differential cross section}

\par
The differential cross sections of inclusive charged hadron production 
as a function of 
\pt {} are measured for an effective mass of the $\Gg$ system
$\Wgg \ge 5 \GeV $,
with a mean value of $\langle \Wgg \rangle \simeq 30 \GeV$,
a photon virtuality $Q^2 \le 8 ~ \GeV ^2 $ and an average photon virtuality
 $\langle Q^2 \rangle \simeq 0.2 ~ \GeV ^2 $. This phase space is defined by cuts at the Monte Carlo generator level. Results are presented in 12 \pt {}
bins  between 0.4 and 20 \GeV .

\par
\begin{sidewaysfigure}
  \begin{tabular}{cc}
    \includegraphics[width=0.45\linewidth]{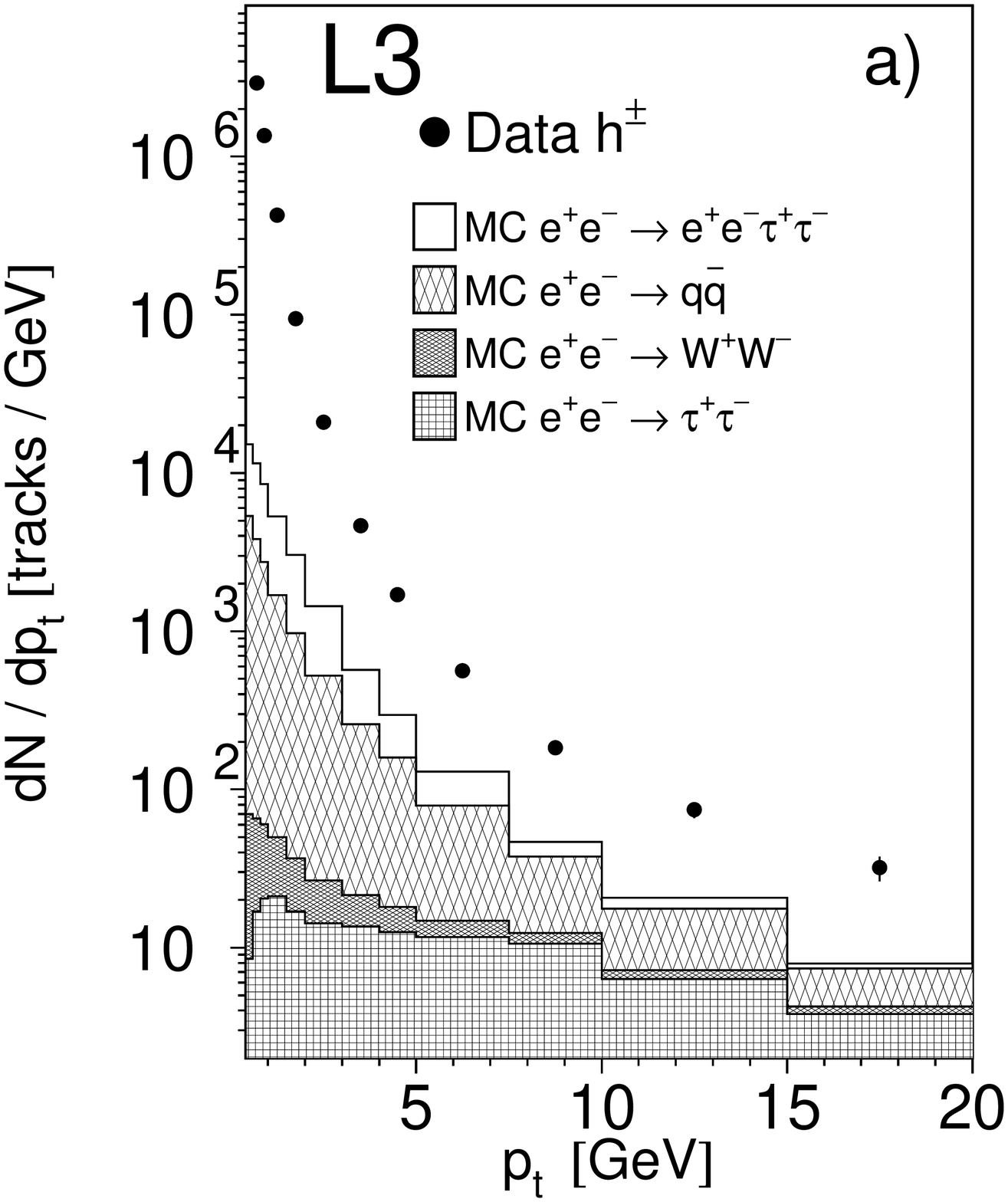}&
    \includegraphics[width=0.45\linewidth]{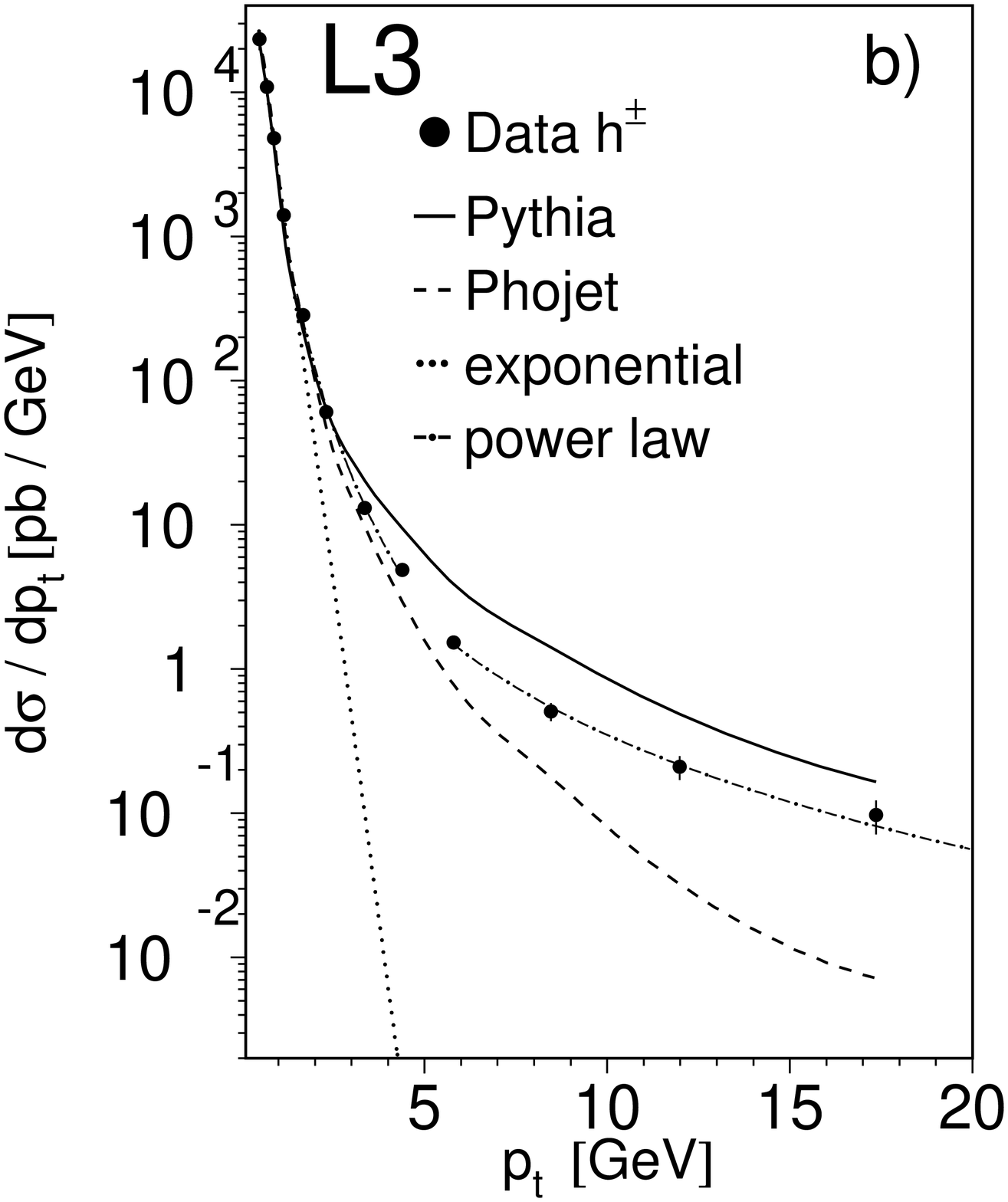}
  \end{tabular}
  \caption{a) Number of selected tracks per \GeV {} in each \pt {} bin and main sources of background. 
   b) Inclusive charged hadron differential cross section \dpt {} 
   fitted with an exponential and power-law functions.
   Monte Carlo predictions are also presented. Statistical and systematic uncertainties are shown. The average $\pt$ value of each bin, $\langle \pt \rangle$, is used.}
  \label{fig:hadrons}
\end{sidewaysfigure}

The distribution of the detected charged hadrons 
in these \pt {} bins is presented
in Figure \ref{fig:hadrons}a. The background
remains very low over the whole \pt {} range.
Events from the \ee \ra {} \ee $\tau^{+} \tau^{-}$  process dominate the background at low \pt {}
while annihilation events dominate it at high \pt .
To measure the cross section, the background is subtracted bin-by-bin
and the data are corrected for the selection efficiency, including acceptance, 
calculated bin-by-bin with PYTHIA.
This selection efficiency
varies from 62\% to 84\%. 
At low \pt , the efficiency decreases due to the effect of the mass and energy cuts. 
At high \pt , it decreases because of the multiplicity cut, since high \pt {}
particles are mainly produced in low multiplicity events.

\par
The level 1 trigger efficiency is obtained
by comparing  the number of events accepted by 
the independent track and   
calorimetric energy\cite{etrig} triggers. It varies from 95\% to 98\%. 
The efficiency of higher level triggers is about 95\% and is measured using
prescaled  events. 
The overall efficiency, taking into account selection and
trigger efficiencies is given in Table \ref{tab:hadrons}.

\par
Sources of systematic uncertainties on the cross section measurements
are the trigger efficiency estimation, the background subtraction, 
the selection procedure and the Monte Carlo modeling. Their contributions are shown in Table \ref{tab:err}.
The uncertainty on the trigger efficiency and on the background subtraction 
are of a statistical nature.
The uncertainty due to the selection procedure is evaluated by 
repeating the analysis with different selection criteria:
the multiplicity cut is moved from 5 to 7 objects, the energy cut is moved to 0.35 \sqs {}
and the number of hits of the tracks is moved to 70\% of that expected.
The sum in quadrature of the differences between these and the reference results
is listed in Table \ref{tab:err}.
Varying other criteria give negligible contributions. 
To evaluate the uncertainty on the Monte Carlo modeling,
the selection efficiency is determined using only one of the PYTHIA subprocesses: VDM-VDM,
direct-direct or resolved-resolved. 
The systematic uncertainty is assigned as the average difference between these values and the
reference Monte Carlo.
The larger contribution comes from the difference
between direct and other processes.

\par
The differential cross section of charged hadron production
as a function of \pt {} is presented in Figure \ref{fig:hadrons}b 
and in Table  \ref{tab:hadrons}.
The migration due to the \pt {} resolution
does not affect these results. 
This was verified by performing
a one-step Bayesian  unfolding\cite{bayes} of the track \pt {} distribution which
give results compatible, within errors, with those obtained using 
the bin-by-bin correction.

\par
The steep decrease of \dpt {} 
in the range $0.4 < \pt < 1.5 \GeV ${} is described by an exponential
 of the form $A \, {\rm exp}({-\pt/\langle\pt\rangle})$
with a mean  value of \mbox{$\langle\pt\rangle \simeq 232$ \MeV}. 
This behaviour is characteristic  of hadrons produced by soft interactions  
and is similar to that obtained in hadron-hadron
and photon-hadron collisions \cite{perl}. 
At higher \pt {} the  differential cross section
is better represented 
by power law functions $A \pt^{-B}$,
as expected by the onset of QCD processes.
For $1.5 \GeV < \pt < 5 \GeV$, $B \simeq 4.2$ and for $5 \GeV < \pt < 20 \GeV$, $B \simeq 2.6$. 
The results of the fits are drawn on  Figure \ref{fig:hadrons}b
where the data are also compared to Monte Carlo predictions.
PYTHIA is slightly above the data, whereas PHOJET is too
low by more than one order of magnitude.
These results are consistent with our findings in inclusive \pz {} production \cite{pz}. 

\section{Charged pions and charged kaons}

\par
Assuming the fragmentation function implemented in JETSET are correct, the \ppm {} and the \kpm {}
inclusive cross sections are extracted from the charged hadron cross section.
Their ratios relative to charged hadrons are estimated bin-by-bin
from Monte Carlo.  Above 5 \GeV , they are almost constant.
Their uncertainty is calculated in the same way as the uncertainty on the Monte Carlo modeling
of the selection efficiency, by
using different subprocesses in PYTHIA. This gives an additional systematic 
uncertainty of from 2\%
to 12\% for pions and from 14\% to 24\% for kaons. 

\par
\begin{sidewaysfigure}
  \begin{tabular}{cc}
    \includegraphics[width=0.45\linewidth]{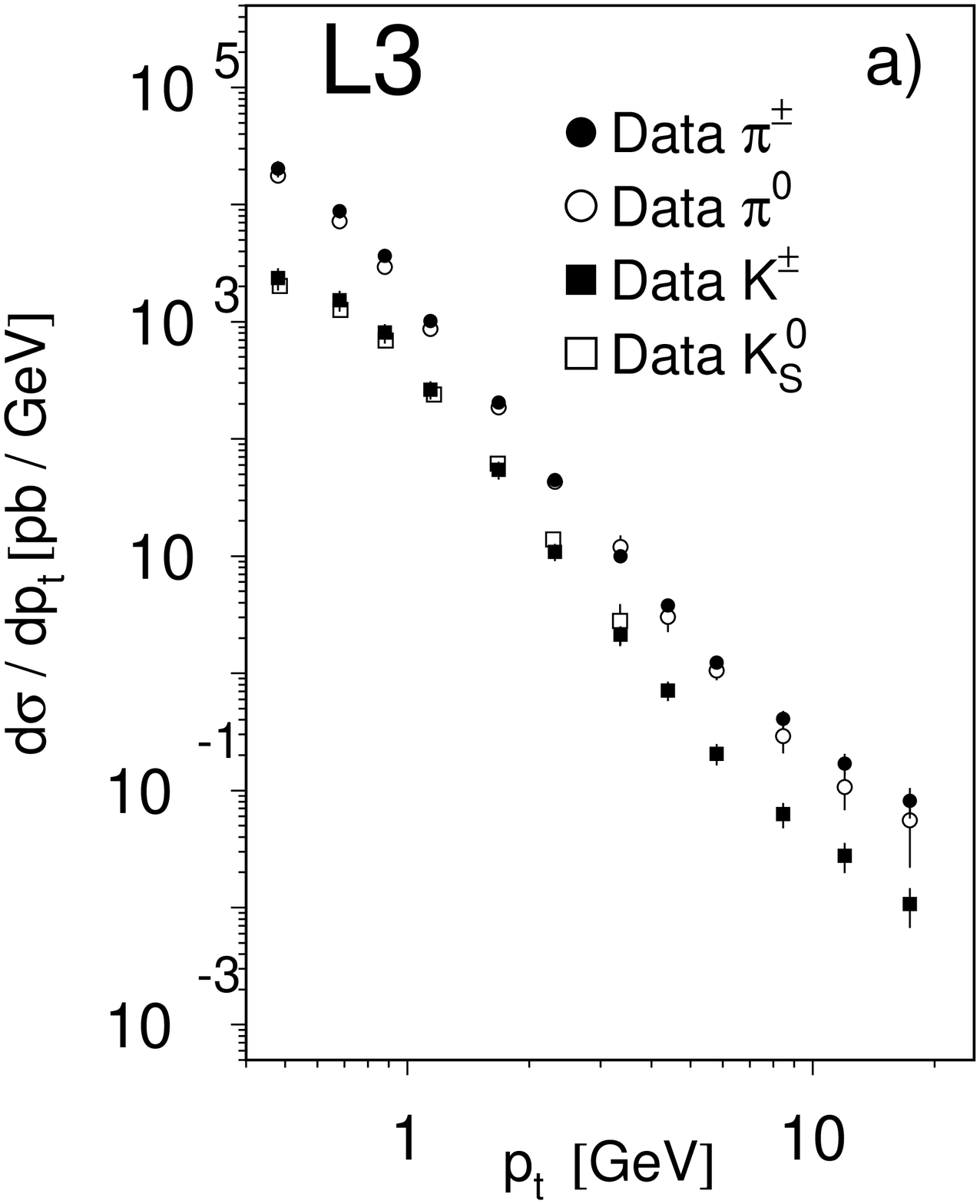}&
    \includegraphics[width=0.45\linewidth]{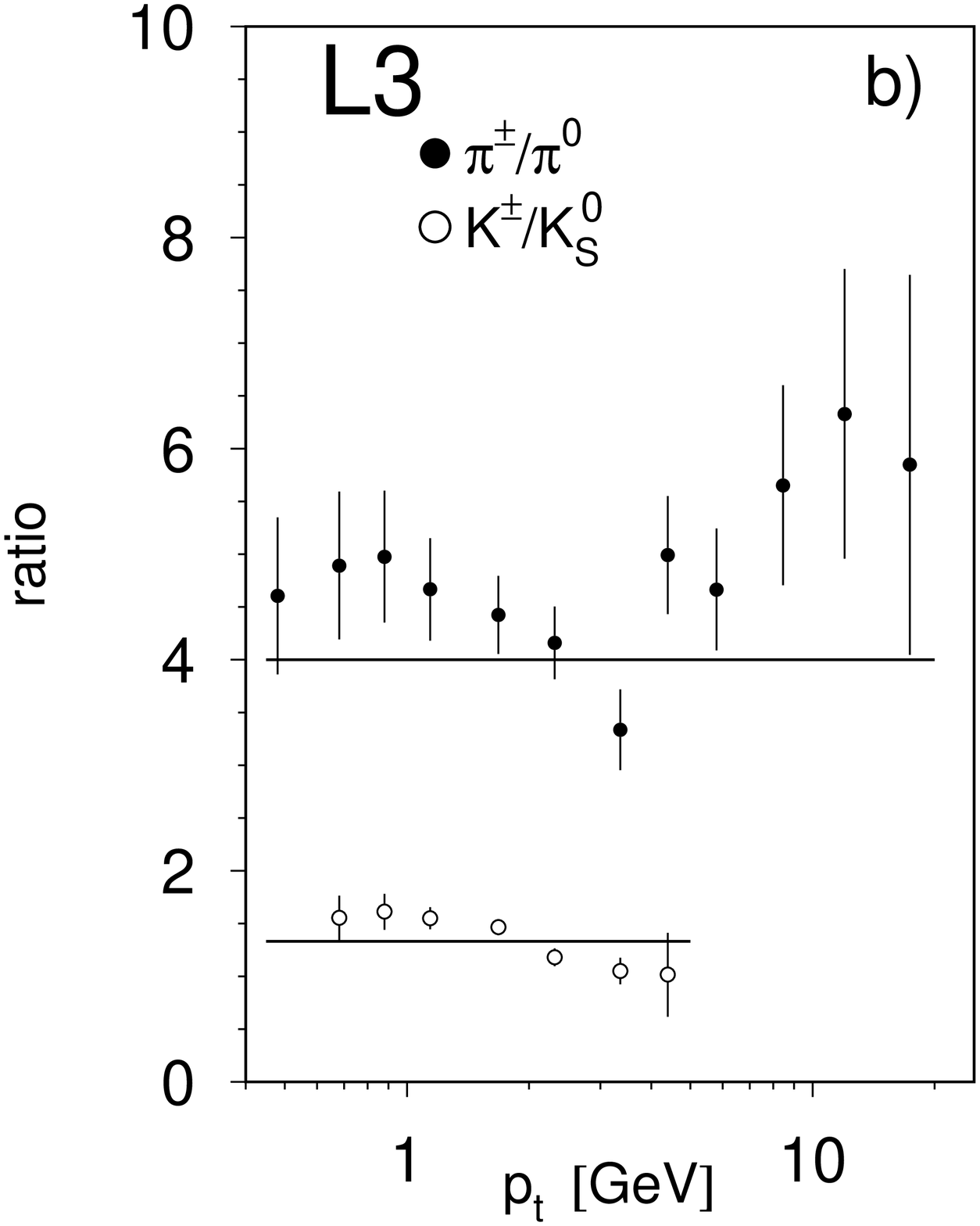}
  \end{tabular}
  \caption{ a) Differential cross section \dpt {} for
  inclusive pion and kaon production.
  The  \pipm {} data are 
  compared to the inclusive \pz {} measurement [3] scaled by a factor 4. The \kpm {} data are
  compared to the inclusive \ks {} measurement [3] scaled by a factor 4/3.
  b) Cross section ratios for pions and kaons. Good agreement is found
  with the expected values (horizontal lines). Statistical and systematic uncertainties are shown. The average $\pt$ value of each bin, $\langle \pt \rangle$, is used.
   }
  \label{fig:kaons}
\end{sidewaysfigure}

The differential cross sections for \ppm {}  and \kpm {} production
as a function of \pt {} are presented in Figure \ref{fig:kaons} 
and in Table  \ref{tab:kaons}.
The \ppm {} data are compared to the previous \pz {} data \cite{pz}  scaled up by a factor 4:
a factor 2 to correct for the $|\eta| < 0.5$ interval used for the \pz {} measurement
and a factor 2 to take into account the isospin symmetry.
A good agreement is found between these two measurements
as shown in Figure \ref{fig:kaons}.
The \kpm {} data are compared to the previous results of \ks {} data \cite{pz}  
scaled up by a factor 4/3:
a factor 2/3 to correct for the $|\eta| < 1.5$ interval of the \ks {} measurement
and a factor 2 to take into account unobserved $\rm K^0_L$ decays. 
Good agreement is found between these two measurements
as shown in Figure \ref{fig:kaons}.
These agreements show a good consistency with the data of the fragmentation functions as implemented in JETSET.

\par
\begin{figure}
    \includegraphics[width=1\linewidth]{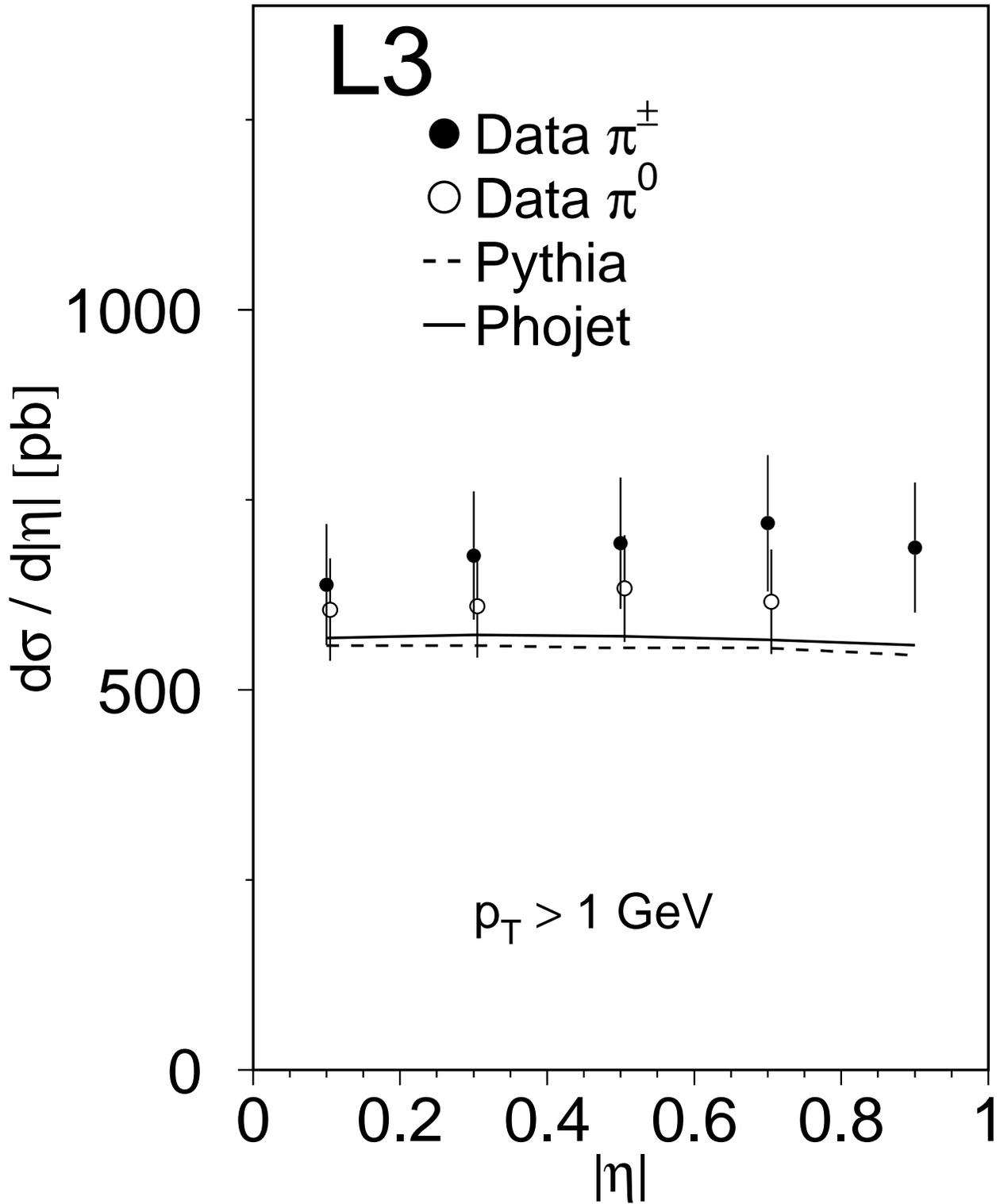}
  \caption{Inclusive \pipm {} differential cross section \deta {} for \pt $> 1$ \GeV {} 
  compared to the inclusive \pz {} measurement [3] scaled by a factor 2 and two Monte Carlo predictions.
   Statistical and systematic uncertainties are shown.}
  \label{fig:eta}
\end{figure}

The differential cross section of \ppm {} production
as a function of $|\eta|$ for $\pt > 1 \GeV$   is shown in Figure \ref{fig:eta} 
and in Table  \ref{tab:eta}.
The cross section is almost constant in this $\eta$ range.  
It agrees well with the \pz {} measurement\cite{pz}. For different \pt {} cuts, Monte Carlo and
QCD predictions describe well the uniform $\eta$ distribution, while the agreement in
the absolute rate
depends on the \pt {} range considered.

\par
\begin{sidewaysfigure}
   \begin{tabular}{cc}
    \includegraphics[width=0.45\linewidth]{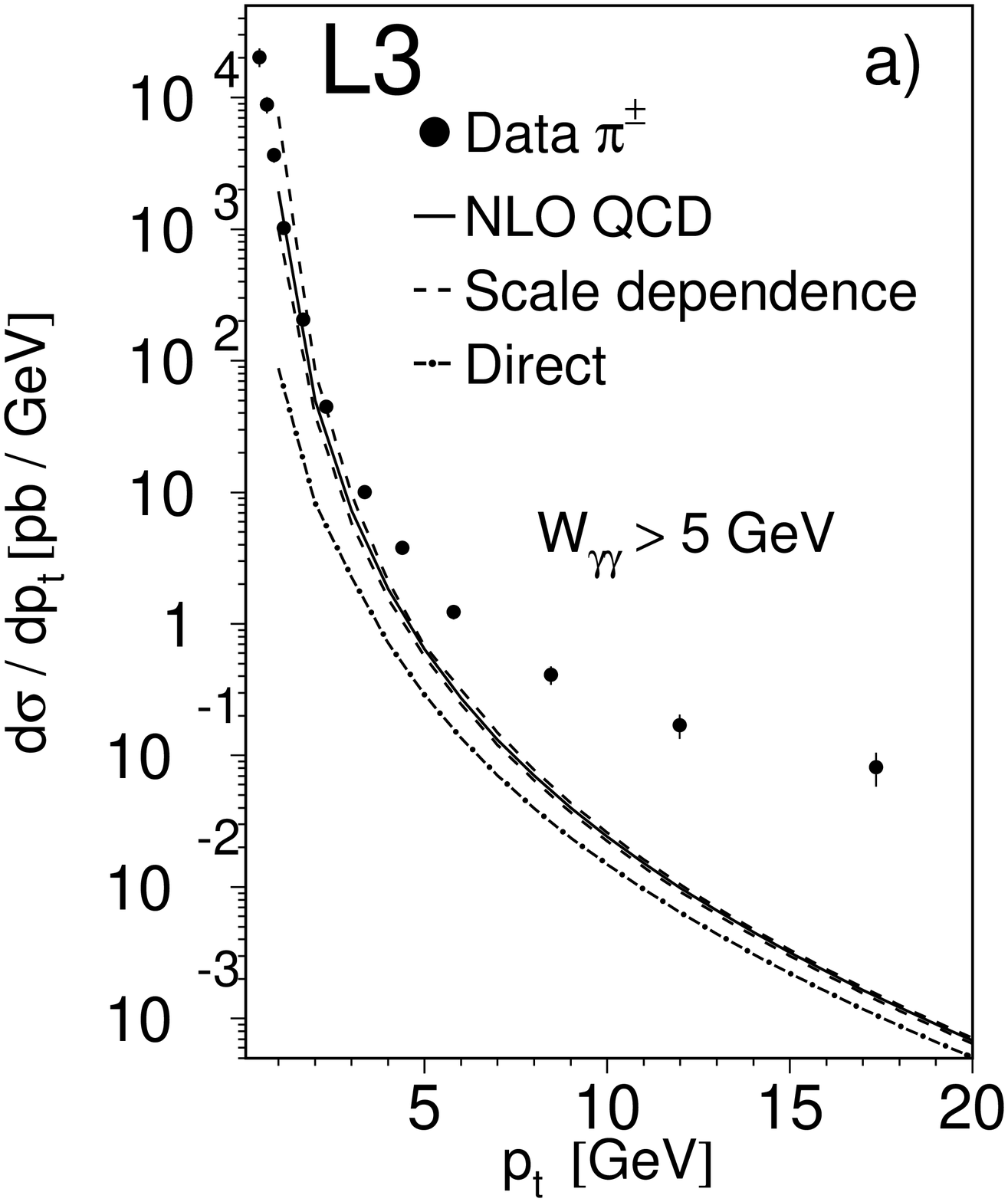} &
    \includegraphics[width=0.45\linewidth]{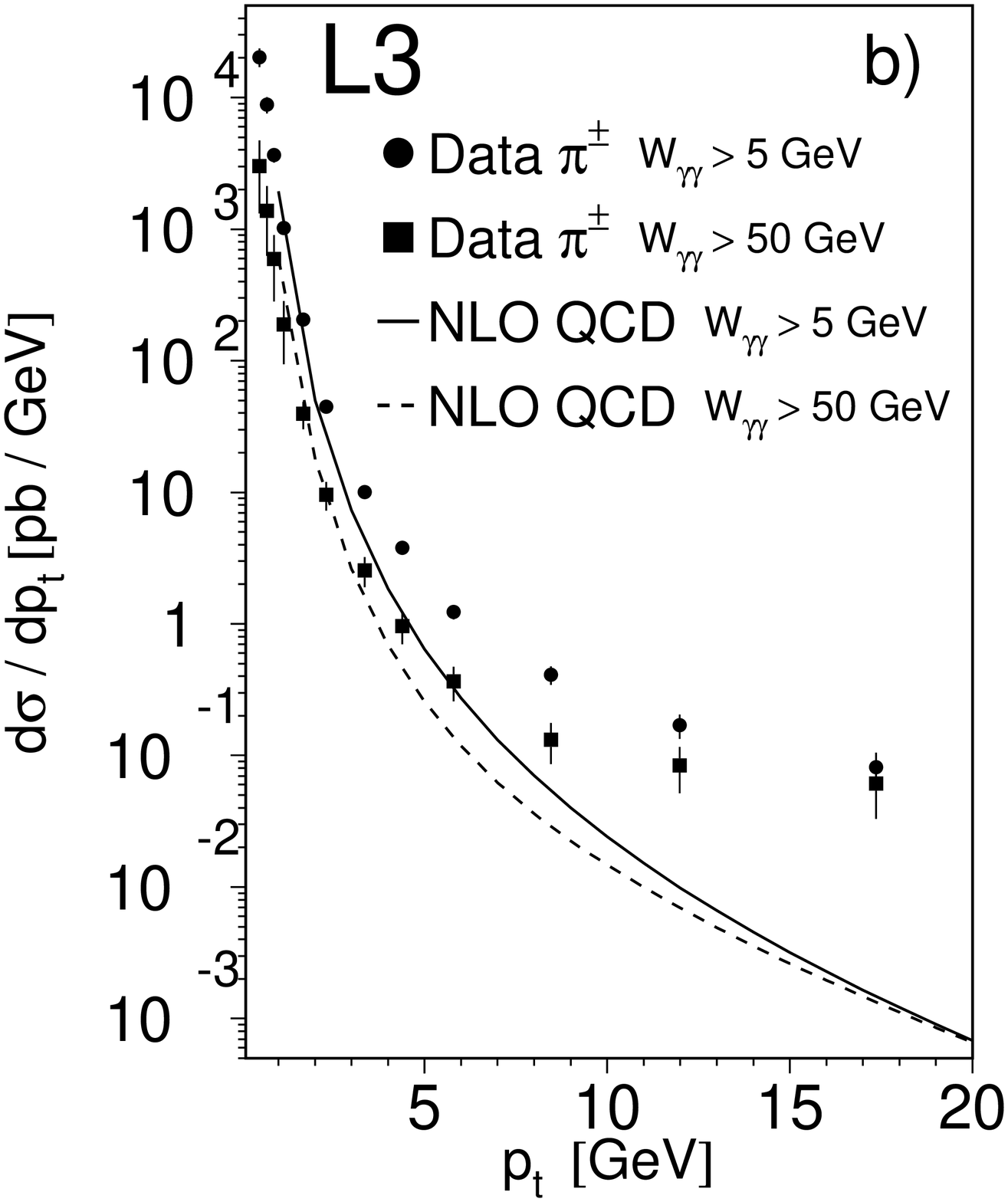}
  \end{tabular}
  \caption{ a) Inclusive \pipm {} differential cross section \dpt {} 
  compared to NLO QCD calculations [19]
  for $\Wgg > 5$ \GeV . 
  The dashed-dotted line corresponds to the direct subprocess. 
  The dashed lines represent the scale uncertainty of the calculations. 
  b) Inclusive \pipm {} differential cross section \dpt {} with different $\Wgg$ cuts.  The average $\pt$ value of each bin, $\langle \pt \rangle$, is used.}
  \label{fig:pions}
\end{sidewaysfigure}

In Figure \ref{fig:pions}a the data are compared
to  analytical NLO QCD predictions \cite{kniehl2, kniehl}. For this calculation,
the flux of quasi-real  photons is obtained using
the Equivalent Photon Approximation \cite{EPA}, taking into account both     
transverse and  longitudinal virtual photons. 
The interacting particles can be point-like photons or 
partons  from the $\gamma \ra \rm{q\bar{q}}$ process, which
evolve into quarks and gluons.
The NLO parton density functions of
Reference  \citen{aurenche} are used and
all elementary $2 \ra 2$ and $2\ra 3$ processes  are considered. 
New NLO fragmentation functions \cite{FF}
are used.
The renormalization, factorisation and fragmentation scales
are taken to be equal: $\mu=M=M_F=\xi \pt$ \cite{kniehl}, with $\xi = 1$ for the central value.
The scale uncertainty in the NLO calculation is estimated by varying the value of $\xi$
from 0.5 to 2.0. 
The agreement with the data is poor in the high-\pt {} range for any choice of scale.

\par
To test NLO QCD calculations in regions where non-perturbative subprocesses are better suppressed, we have also measured
differential cross sections of \ppm {} production for  $\Wgg >$ 10, 30 and 50 \GeV . The results are shown in
Table \ref{tab:wgg} and Figure \ref{fig:pions}b. 
The discrepancy between the calculations and data at high \pt {} is not significantly 
reduced by these stringent more $\Wgg$ cuts.

\par
Similar calculations were previously compared to $\gamma$p reactions at HERA up to a $\pt$ of $12\GeV$ and to ${\rm\bar{p}}$p collisions up to a $\pt$ of $20\GeV$. Good agreement was found~\cite{kkp}. In the $\gamma\gamma$ channel, an excess of data with respect to NLO QCD was observed in tagged events at PETRA experiments~\cite{kniehl}. No discrepancy is observed with the OPAL data which explore a $\pt$ range up to $10\GeV$. In this range, our data and the OPAL ones are well in agreement within the quoted uncertainties. A discrepancy with NLO QCD is revealed by our data which extend the measurement to higher  $\pt$ values.

\newpage
\section*{Author List}
\input namelist261.tex

\begin{table}
  \begin{center}

    \begin{tabular}{|r@{~$-$}l|c|r@{~$\pm$~}l|r@{~$\pm$~}r@{~$\pm$~}l@{~}l|}
    \hline
    \multicolumn{2}{|c|}{\pt}  & $\langle\pt\rangle$ &  \multicolumn{2}{|c|}{Efficiency}
     & \multicolumn{4}{|c|}{\dpt} \\
     \multicolumn{2}{|c|}{[\GeV]}  & [\GeV]& \multicolumn{2}{|c|}{[\%] } 
     & \multicolumn{4}{|c|}{ [pb/\GeV] }\\
    \hline
       \phantom{0}0.4 & \phantom{0}0.6 & \phantom{0}0.48 & 62.4 & \phantom{0}7.7 & (23.4 & 0.1 & \phantom{0}3.7)&$\times 10 ^3$\\
       \phantom{0}0.6 & \phantom{0}0.8 & \phantom{0}0.68 & 64.5 & \phantom{0}6.9 & (10.9 & 0.1 & \phantom{0}1.5)&$\times 10 ^3$\\
       \phantom{0}0.8 & \phantom{0}1.0 & \phantom{0}0.88 & 67.7 & \phantom{0}6.0 & (48.0 & 0.1 & \phantom{0}5.9)&$\times 10 ^2$\\
       \phantom{0}1.0 & \phantom{0}1.5 & \phantom{0}1.14 & 72.4 & \phantom{0}4.8 & (14.1 & 0.1 & \phantom{0}1.4)&$\times 10 ^2$\\
       \phantom{0}1.5 & \phantom{0}2.0 & \phantom{0}1.68 & 77.4 & \phantom{0}3.7 & (28.5 & 0.1 & \phantom{0}2.2)&$\times 10$\\
       \phantom{0}2.0 & \phantom{0}3.0 & \phantom{0}2.31 & 77.2 & \phantom{0}4.3 & (60.9 & 0.5 & \phantom{0}4.4)& \\
       \phantom{0}3.0 & \phantom{0}4.0 & \phantom{0}3.36 & 75.0 & \phantom{0}5.2 & (13.1 & 0.2 & \phantom{0}1.0)& \\
       \phantom{0}4.0 & \phantom{0}5.0 & \phantom{0}4.39 & 69.5 & \phantom{0}5.8 & (48.7 & 1.3 & \phantom{0}4.2)&$\times 10 ^{-1}$\\
       \phantom{0}5.0 & \phantom{0}7.5 & \phantom{0}5.79 & 68.1 & \phantom{0}6.8 & (15.3 & 0.4 & \phantom{0}1.6)&$\times 10 ^{-1}$\\
       \phantom{0}7.5 & \phantom{}10.0 & \phantom{0}8.46 & 65.2 & \phantom{0}8.7 & (50.9 & 2.5 & \phantom{0}7.0)&$\times 10 ^{-2}$\\
       \phantom{}10.0 & \phantom{}15.0 & \phantom{}11.98 & 61.9 & \phantom{}11.0 & (21.0 & 1.2 & \phantom{0}3.8)&$\times 10 ^{-2}$\\
       \phantom{}15.0 & \phantom{}20.0 & \phantom{}17.36 & 59.8 & \phantom{}14.7 & (97.1 & 8.4 & \phantom{}24.3)&$\times 10 ^{-3}$\\
     \hline
   \end{tabular}

   \caption{Transverse momentum range and average value from the data with the corresponding
   overall efficiency and differential cross section for inclusive charged hadron production
   for $\Wgg >$ 5 \GeV {} and $|\eta| < 1$.
   The uncertainty on the efficiency is systematic.
   The first uncertainty on the cross section is statistical and the second systematic.}
    \label{tab:hadrons}

  \end{center}
\end{table}

\begin{table}
  \begin{center}

    \begin{tabular}{|r@{~$-$}l||c|c|c|c|}
    \hline
    \multicolumn{2}{|c||}{\pt     }  & trigger & background & selection & Monte Carlo \\
    \multicolumn{2}{|c||}{ [\GeV] }  & efficiency [\%] & subtraction [\%] & procedure [\%] & modeling [\%]  \\
    \hline
       \phantom{0}0.4 & \phantom{0}0.6 & 0.1 &       $<$ 0.1 & \phantom{}10.1 & \phantom{}12.4 \\
       \phantom{0}0.6 & \phantom{0}0.8 & 0.1 &       $<$ 0.1 & \phantom{0}9.2 & \phantom{}10.6 \\
       \phantom{0}0.8 & \phantom{0}1.0 & 0.2 &       $<$ 0.1 & \phantom{0}8.4 & \phantom{0}8.9 \\
       \phantom{0}1.0 & \phantom{0}1.5 & 0.2 &       $<$ 0.1 & \phantom{0}7.5 & \phantom{0}6.6 \\
       \phantom{0}1.5 & \phantom{0}2.0 & 0.4 &       $<$ 0.1 & \phantom{0}5.9 & \phantom{0}4.8 \\
       \phantom{0}2.0 & \phantom{0}3.0 & 0.5 & \phantom{$<$} 0.1 & \phantom{0}4.5 & \phantom{0}5.6 \\
       \phantom{0}3.0 & \phantom{0}4.0 & 0.9 & \phantom{$<$} 0.4 & \phantom{0}3.1 & \phantom{0}6.9 \\
       \phantom{0}4.0 & \phantom{0}5.0 & 1.2 & \phantom{$<$} 0.9 & \phantom{0}2.3 & \phantom{0}8.2 \\
       \phantom{0}5.0 & \phantom{0}7.5 & 1.2 & \phantom{$<$} 1.3 & \phantom{0}1.6 & \phantom{}10.0 \\
       \phantom{0}7.5 & \phantom{}10.0 & 1.2 & \phantom{$<$} 2.5 & \phantom{0}1.2 & \phantom{}13.3 \\
       \phantom{}10.0 & \phantom{}15.0 & 1.2 & \phantom{$<$} 3.1 & \phantom{0}1.1 & \phantom{}17.8 \\
       \phantom{}15.0 & \phantom{}20.0 & 1.2 & \phantom{$<$} 4.2 & \phantom{0}1.1 & \phantom{}24.6 \\
     \hline
    \end{tabular}

    \caption{Systematic uncertainty on the charged hadron cross section
    due to trigger
    efficiency,  background subtraction,
    selection procedure and Monte Carlo modeling.}
    \label{tab:err}

  \end{center}
\end{table}

\begin{table}
  \begin{center}

    \begin{tabular}{|c||r@{~$\pm$~}r@{~$\pm$~}l@{~}l|r@{~$\pm$~}r@{~$\pm$~}l@{~}l|}
    \hline
    $\langle\pt\rangle$ & \multicolumn{4}{|c|}{\dpt {} for pions} & \multicolumn{4}{|c|}{\dpt {} for kaons} \\ 
    \phantom{0}[\GeV]\phantom{0}& \multicolumn{4}{|c|}{ [pb/\GeV] }& \multicolumn{4}{|c|}{ [pb/\GeV] }\\ 
    \hline
       \phantom{0}0.48 & (20.3 & 0.1 & \phantom{0}3.3)&$\times 10 ^3$    & (23.7 & 0.1 & \phantom{0}5.1)&$\times 10 ^2$\\ 
       \phantom{0}0.68 & (88.1 & 0.1 & \phantom{}12.5)&$\times 10 ^2$    & (15.3 & 0.1 & \phantom{0}3.1)&$\times 10 ^2$\\ 
       \phantom{0}0.88 & (36.5 & 0.1 & \phantom{0}4.5)&$\times 10 ^2$    & (80.7 & 0.2 & \phantom{}15.3)&$\times 10$\\ 
       \phantom{0}1.14 & (10.2 & 0.1 & \phantom{0}1.0)&$\times 10 ^2$    & (26.4 & 0.1 & \phantom{0}4.7)&$\times 10$\\ 
       \phantom{0}1.68 & (20.5 & 0.1 & \phantom{0}1.6)&$\times 10$       & (54.3 & 0.4 & \phantom{0}9.2)&                  \\ 
       \phantom{0}2.31 & (44.7 & 0.3 & \phantom{0}3.5)&                 & (10.9 & 0.1 & \phantom{0}1.9)&                  \\
       \phantom{0}3.36 & (10.0 & 0.2 & \phantom{0}0.9)&                 & (21.3 & 0.4 & \phantom{0}3.8)&$\times 10 ^{-1}$\\
       \phantom{0}4.39 & (37.8 & 1.0 & \phantom{0}3.6)&$\times 10 ^{-1}$ & (71.3 & 2.4 & \phantom{}13.4)&$\times 10 ^{-2}$\\ 
       \phantom{0}5.79 & (12.3 & 0.4 & \phantom{0}1.4)&$\times 10 ^{-1}$ & (20.6 & 0.8 & \phantom{0}4.2)&$\times 10 ^{-2}$\\ 
       \phantom{0}8.46 & (41.0 & 2.1 & \phantom{0}6.2)&$\times 10 ^{-2}$ & (62.9 & 4.1 & \phantom{}14.8)&$\times 10 ^{-3}$\\
       \phantom{}11.98 & (16.9 & 1.0 & \phantom{0}3.4)&$\times 10 ^{-2}$ & (27.7 & 2.1 & \phantom{0}7.7)&$\times 10 ^{-3}$\\ 
       \phantom{}17.36 & (81.3 & 7.1 & \phantom{}22.5)&$\times 10 ^{-3}$ & (10.7 & 1.4 & \phantom{0}3.7)&$\times 10 ^{-3}$\\ 
     \hline
   
   \end{tabular}

    \caption{Differential cross section as a function of \pt {} for inclusive 
    \ppm {} and \kpm {} production for $\Wgg >$ 5 \GeV {} and $|\eta| < 1$.
   The first uncertainty on the cross section is statistical and the second systematic.}
    \label{tab:kaons}

  \end{center}
\end{table}

\begin{table}
  \begin{center}

    \begin{tabular}{|c||r@{~$\pm$~}r@{~$\pm$~}l@{~}l|r@{~$\pm$~}r@{~$\pm$~}l@{~}l|r@{~$\pm$~}r@{~$\pm$~}l@{~}l|}
    \hline
    $\langle\pt\rangle$    & \multicolumn{4}{|c|}{\dpt {} [pb/\GeV]} & \multicolumn{4}{|c|}{\dpt {} [pb/\GeV]} & \multicolumn{4}{|c|}{\dpt {} [pb/\GeV]} \\
    \phantom{0}[\GeV]\phantom{0}& \multicolumn{4}{|c|}{$\Wgg > 10$ \GeV}  & \multicolumn{4}{|c|}{$\Wgg > 30$ \GeV}  & \multicolumn{4}{|c|}{$\Wgg > 50$ \GeV}\\
    \hline
       \phantom{0}0.48 & (13.7 & 0.1 & \phantom{0}3.2)&$\times 10 ^3$    & (56.2 & 0.2 & \phantom{}23.8)&$\times 10 ^2$    & (30.1 & 0.2 & \phantom{}17.0)&$\times 10 ^2$\\
       \phantom{0}0.68 & (60.6 & 0.1 & \phantom{}12.5)&$\times 10 ^2$    & (25.3 & 0.1 & \phantom{0}9.9)&$\times 10 ^2$    & (13.8 & 0.1 & \phantom{0}7.5)&$\times 10 ^2$\\
       \phantom{0}0.88 & (25.7 & 0.1 & \phantom{0}4.5)&$\times 10 ^2$    & (10.9 & 0.1 & \phantom{0}3.9)&$\times 10 ^2$    & (59.3 & 0.6 & \phantom{}31.2)&$\times 10$\\
       \phantom{0}1.14 & (74.3 & 0.2 & \phantom{}10.3)&$\times 10$       & (33.5 & 0.2 & \phantom{}10.7)&$\times 10$       & (18.9 & 0.2 & \phantom{0}9.5)&$\times 10$\\
       \phantom{0}1.68 & (15.6 & 0.1 & \phantom{0}1.0)&$\times 10$       & (73.5 & 0.8 & \phantom{0}8.9)&                  & (39.4 & 0.7 & \phantom{0}9.4)&          \\
       \phantom{0}2.31 & (36.4 & 0.3 & \phantom{0}2.2)&                  & (17.2 & 0.3& \phantom{0}2.0)&                  & (96.1 & 2.3 & \phantom{}23.4)&$\times 10 ^{-1}$\\
       \phantom{0}3.36 & (88.8 & 1.5 & \phantom{0}5.8)&$\times 10 ^{-1}$ & (40.5 & 1.1 & \phantom{0}4.6)&$\times 10 ^{-1}$ & (25.5 & 1.1 & \phantom{0}6.5)&$\times 10 ^{-1}$\\
       \phantom{0}4.39 & (35.7 & 1.0 & \phantom{0}2.9)&$\times 10 ^{-1}$ & (18.9 & 0.8 & \phantom{0}2.2)&$\times 10 ^{-1}$ & (96.5 & 6.0 & \phantom{}26.0)&$\times 10 ^{-2}$\\
       \phantom{0}5.79 & (11.7 & 0.4 & \phantom{0}1.2)&$\times 10 ^{-1}$ & (60.4 & 2.5 & \phantom{0}7.2)&$\times 10 ^{-2}$ & (36.5 & 2.1 & \phantom{}10.5)&$\times 10 ^{-2}$\\
       \phantom{0}8.46 & (37.8 & 2.0 & \phantom{0}5.6)&$\times 10 ^{-2}$ & (22.1 & 1.5 & \phantom{0}2.9)&$\times 10 ^{-2}$ & (13.1 & 1.2 & \phantom{0}4.4)&$\times 10 ^{-2}$\\
       \phantom{}11.98 & (16.4 & 1.0 & \phantom{0}3.4)&$\times 10 ^{-2}$ & (12.7 & 0.9 & \phantom{0}1.8)&$\times 10 ^{-2}$ & (84.1 & 7.4 & \phantom{}31.5)&$\times 10 ^{-3}$\\
       \phantom{}17.36 & (78.9 & 7.0 & \phantom{}23.9)&$\times 10 ^{-3}$ & (60.0 & 6.3 & \phantom{}10.1)&$\times 10 ^{-3}$ & (61.3 & 7.5 & \phantom{}27.2)&$\times 10 ^{-3}$\\
     \hline
   \end{tabular}
    \caption{Differential cross section as a function of \pt {} for inclusive \ppm {}
    production for $|\eta| < 1$ and different $\Wgg$ cuts.
   The first uncertainty on the cross section is statistical and the second systematic.}
    \label{tab:wgg}
         
  \end{center}
\end{table}

\begin{table}
  \begin{center}

    \begin{tabular}{|r@{~$-$}l||r@{~$\pm$~}r@{~$\pm$~}l|}
    \hline
    \multicolumn{2}{|c||}{$|\eta |$}  & \multicolumn{3}{|c|}{\deta {} [pb]} \\
    \hline
       \phantom{0}0.0 & \phantom{0}0.2 & 638 & 3 & 80 \\
       \phantom{0}0.2 & \phantom{0}0.4 & 677 & 3 & 84 \\
       \phantom{0}0.4 & \phantom{0}0.6 & 693 & 4 & 86 \\
       \phantom{0}0.6 & \phantom{0}0.8 & 719 & 4 & 90 \\
       \phantom{0}0.8 & \phantom{0}1.0 & 687 & 4 & 86 \\
     \hline
   \end{tabular}

   \caption{Differential cross section 
    as a function of  $|\eta |$ for inclusive \ppm {} production for $\Wgg >$ 5 \GeV {} and $\pt > 1$ \GeV .
   The first uncertainty on the cross section is statistical and the second systematic.}
    \label{tab:eta}

  \end{center}
\end{table}

\newpage

\end{document}

%% file: namelist261.tex
\typeout{   }     
\typeout{Using author list for paper 261 -  }
\typeout{$Modified: Jul 15 2001 by smele $}
\typeout{!!!!  This should only be used with document option a4p!!!!}
\typeout{   }
%
%
%
%
%
%

\newcount\tutecount  \tutecount=0
\def\tutenum#1{\global\advance\tutecount by 1 \xdef#1{\the\tutecount}}
\def\tute#1{$^{#1}$}
\tutenum\aachen            
\tutenum\nikhef            
\tutenum\mich              
\tutenum\lapp              
\tutenum\basel             
\tutenum\lsu               
\tutenum\beijing           
\tutenum\bologna           
\tutenum\tata              
\tutenum\ne                
\tutenum\bucharest         
\tutenum\budapest          
\tutenum\mit               
\tutenum\panjab            
\tutenum\debrecen          
\tutenum\dublin            
\tutenum\florence          
\tutenum\cern              
\tutenum\wl                
\tutenum\geneva            
\tutenum\hefei             
\tutenum\lausanne          
\tutenum\lyon              
\tutenum\madrid            
\tutenum\florida           
\tutenum\milan             
\tutenum\moscow            
\tutenum\naples            
\tutenum\cyprus            
\tutenum\nymegen           
\tutenum\caltech           
\tutenum\perugia           
\tutenum\peters            
\tutenum\cmu               
\tutenum\potenza           
\tutenum\prince            
\tutenum\riverside         
\tutenum\rome              
\tutenum\salerno           
\tutenum\ucsd              
\tutenum\sofia             
\tutenum\korea             
\tutenum\purdue            
\tutenum\psinst            
\tutenum\zeuthen           
\tutenum\eth               
\tutenum\hamburg           
\tutenum\taiwan            
\tutenum\tsinghua          

{
\parskip=0pt
\noindent
{\bf The L3 Collaboration:}
\ifx\selectfont\undefined
 \baselineskip=10.8pt
 \baselineskip\baselinestretch\baselineskip
 \normalbaselineskip\baselineskip
 \ixpt
\else
 \fontsize{9}{10.8pt}\selectfont
\fi
\medskip
\tolerance=10000
\hbadness=5000
\raggedright
\hsize=162truemm\hoffset=0mm
\def\r{\rlap,}
\noindent

P.Achard\r\tute\geneva\ 
O.Adriani\r\tute{\florence}\ 
M.Aguilar-Benitez\r\tute\madrid\ 
J.Alcaraz\r\tute{\madrid,\cern}\ 
G.Alemanni\r\tute\lausanne\
J.Allaby\r\tute\cern\
A.Aloisio\r\tute\naples\ 
M.G.Alviggi\r\tute\naples\
H.Anderhub\r\tute\eth\ 
V.P.Andreev\r\tute{\lsu,\peters}\
% ...
\newpage
\begin{list}{A}{\itemsep=0pt plus 0pt minus 0pt\parsep=0pt plus 0pt minus 0pt
                \topsep=0pt plus 0pt minus 0pt}
\item[\aachen]
 III. Physikalisches Institut, RWTH, D-52056 Aachen, Germany$^{\S}$
\item[\nikhef] National Institute for High Energy Physics, NIKHEF, 
     and University of Amsterdam, NL-1009 DB Amsterdam, The Netherlands
\item[\mich] University of Michigan, Ann Arbor, MI 48109, USA
\item[\lapp] Laboratoire d'Annecy-le-Vieux de Physique des Particules, 
     LAPP,IN2P3-CNRS, BP 110, F-74941 Annecy-le-Vieux CEDEX, France
\item[\basel] Institute of Physics, University of Basel, CH-4056 Basel,
     Switzerland
\item[\lsu] Louisiana State University, Baton Rouge, LA 70803, USA
\item[\beijing] Institute of High Energy Physics, IHEP, 
  100039 Beijing, China$^{\triangle}$ 
\item[\bologna] University of Bologna and INFN-Sezione di Bologna, 
     I-40126 Bologna, Italy
\item[\tata] Tata Institute of Fundamental Research, Mumbai (Bombay) 400 005, India
\item[\ne] Northeastern University, Boston, MA 02115, USA
\item[\bucharest] Institute of Atomic Physics and University of Bucharest,
     R-76900 Bucharest, Romania
\item[\budapest] Central Research Institute for Physics of the 
     Hungarian Academy of Sciences, H-1525 Budapest 114, Hungary$^{\ddag}$
\item[\mit] Massachusetts Institute of Technology, Cambridge, MA 02139, USA
\item[\panjab] Panjab University, Chandigarh 160 014, India.
\item[\debrecen] KLTE-ATOMKI, H-4010 Debrecen, Hungary$^\P$
\item[\dublin] Department of Experimental Physics,
  University College Dublin, Belfield, Dublin 4, Ireland
\item[\florence] INFN Sezione di Firenze and University of Florence, 
     I-50125 Florence, Italy
\item[\cern] European Laboratory for Particle Physics, CERN, 
     CH-1211 Geneva 23, Switzerland
\item[\wl] World Laboratory, FBLJA  Project, CH-1211 Geneva 23, Switzerland
\item[\geneva] University of Geneva, CH-1211 Geneva 4, Switzerland
\item[\hefei] Chinese University of Science and Technology, USTC,
      Hefei, Anhui 230 029, China$^{\triangle}$
\item[\lausanne] University of Lausanne, CH-1015 Lausanne, Switzerland
\item[\lyon] Institut de Physique Nucl\'eaire de Lyon, 
     IN2P3-CNRS,Universit\'e Claude Bernard, 
     F-69622 Villeurbanne, France
\item[\madrid] Centro de Investigaciones Energ{\'e}ticas, 
     Medioambientales y Tecnol\'ogicas, CIEMAT, E-28040 Madrid,
     Spain${\flat}$ 
\item[\florida] Florida Institute of Technology, Melbourne, FL 32901, USA
\item[\milan] INFN-Sezione di Milano, I-20133 Milan, Italy
\item[\moscow] Institute of Theoretical and Experimental Physics, ITEP, 
     Moscow, Russia
\item[\naples] INFN-Sezione di Napoli and University of Naples, 
     I-80125 Naples, Italy
\item[\cyprus] Department of Physics, University of Cyprus,
     Nicosia, Cyprus
\item[\nymegen] University of Nijmegen and NIKHEF, 
     NL-6525 ED Nijmegen, The Netherlands
\item[\caltech] California Institute of Technology, Pasadena, CA 91125, USA
\item[\perugia] INFN-Sezione di Perugia and Universit\`a Degli 
     Studi di Perugia, I-06100 Perugia, Italy   
\item[\peters] Nuclear Physics Institute, St. Petersburg, Russia
\item[\cmu] Carnegie Mellon University, Pittsburgh, PA 15213, USA
\item[\potenza] INFN-Sezione di Napoli and University of Potenza, 
     I-85100 Potenza, Italy
\item[\prince] Princeton University, Princeton, NJ 08544, USA
\item[\riverside] University of Californa, Riverside, CA 92521, USA
\item[\rome] INFN-Sezione di Roma and University of Rome, ``La Sapienza",
     I-00185 Rome, Italy
\item[\salerno] University and INFN, Salerno, I-84100 Salerno, Italy
\item[\ucsd] University of California, San Diego, CA 92093, USA
\item[\sofia] Bulgarian Academy of Sciences, Central Lab.~of 
     Mechatronics and Instrumentation, BU-1113 Sofia, Bulgaria
\item[\korea]  The Center for High Energy Physics, 
     Kyungpook National University, 702-701 Taegu, Republic of Korea
\item[\purdue] Purdue University, West Lafayette, IN 47907, USA
\item[\psinst] Paul Scherrer Institut, PSI, CH-5232 Villigen, Switzerland
\item[\zeuthen] DESY, D-15738 Zeuthen, Germany
\item[\eth] Eidgen\"ossische Technische Hochschule, ETH Z\"urich,
     CH-8093 Z\"urich, Switzerland
\item[\hamburg] University of Hamburg, D-22761 Hamburg, Germany
\item[\taiwan] National Central University, Chung-Li, Taiwan, China
\item[\tsinghua] Department of Physics, National Tsing Hua University,
      Taiwan, China
\item[\S]  Supported by the German Bundesministerium 
        f\"ur Bildung, Wissenschaft, Forschung und Technologie
\item[\ddag] Supported by the Hungarian OTKA fund under contract
numbers T019181, F023259 and T037350.
\item[\P] Also supported by the Hungarian OTKA fund under contract
  number T026178.
\item[$\flat$] Supported also by the Comisi\'on Interministerial de Ciencia y 
        Tecnolog{\'\i}a.
\item[$\sharp$] Also supported by CONICET and Universidad Nacional de La Plata,
        CC 67, 1900 La Plata, Argentina.
\item[$\triangle$] Supported by the National Natural Science
  Foundation of China.
\end{list}
}
\vfill
